\documentclass[10pt,twocolumn]{article}
\usepackage[margin=1.9cm]{geometry}
\usepackage{amsmath,amssymb,graphicx,bm,booktabs}
\usepackage[hidelinks]{hyperref}
\usepackage[font=small,labelfont=bf]{caption}
\newtheorem{proposition}{Proposition}
\newcommand{\kB}{k_{\mathrm B}}
\newcommand{\dd}{\mathrm{d}}
\title{\bfseries Energy-fed commitment: a minimal model of random selection followed by reinforcement}
\author{Haidar Issa Abbas\\[2pt]
\small Programa de P\'os-Gradua\c{c}\~ao em Ci\^encia e Engenharia de Materiais,\\
\small Universidade Federal de Campina Grande, Campina Grande, PB, Brazil\\
\small \href{mailto:haidar2brazil@gmail.com}{haidar2brazil@gmail.com} \quad ORCID: \href{https://orcid.org/0009-0003-4043-5025}{0009-0003-4043-5025}}
\date{}
\begin{document}
\twocolumn[
\maketitle
\begin{abstract}
When a symmetric instability is crossed, fluctuations choose one of several equivalent branches. We study a minimal stochastic model of what can then make that choice progressively harder to undo. After selection, the stiffness of the selected state grows in proportion to the energy released each time the structure recovers from a perturbation; in this conservative model that energy equals, up to small and explicitly computed terms, the work absorbed from the perturbing agent. The model is a phenomenological closure with one coupling constant. That energy reinforces a branch already chosen, rather than choosing it, is an assumption of the model, which we state and do not derive. Its consequences are quantitative. The growth of commitment depends on how the structure is driven: geometric under displacement control, linear under energy control and sublinear under force control, so that committed states absorb more work per cycle under displacement control and less under force control. Reinforcement suppresses reversal exponentially; beyond a finite commitment the alternative minimum disappears and reversals become rare crossings of an irreversibility threshold. With unbounded reinforcement a choice is retained forever with positive probability under stated conditions; with saturating reinforcement retention is finite, with a lifetime that grows exponentially with the saturation level. Exact first-passage calculations agree with stochastic simulations; these verify the closure and its analysis but cannot test the physical postulate. We identify the experiments that could: protocols that separate absorbed energy from the number of perturbations, and driven steady states with nonzero housekeeping heat, in which heat and absorbed work differ.
\end{abstract}
\vspace{1em}]

\section{Introduction}
Open systems driven far from equilibrium form structures through instabilities: convection cells, chemical waves, crystal morphologies, chiral domains~\cite{Prigogine,CrossHohenberg}. At a symmetric bifurcation with winner-take-all competition only one of the possible branches is realized, and which one is decided largely by chance~\cite{Kondepudi,KondepudiDeBari,DeBari}; the classic demonstration is the random chirality of sodium chlorate crystals~\cite{Kondepudi}. A structure formed in this way may carry traces of its history~\cite{Prigogine,PrigogineScience}, and a transient perturbation can move a driven system to a different persistent state~\cite{Michaelian}. This paper asks a narrower question: what simple mechanism could make a randomly selected state progressively harder to reverse, and with what statistics?

Several lines of work bear on it. Fluctuation relations for driven systems connect the relative likelihood of outcomes to the heat dissipated in reaching them and to their probability of reverting~\cite{England,Perunov}. Numerical studies of driven multistable networks find that the configurations reached are those that absorb little work from the specific drive~\cite{Kedia,GoldEngland}, and low-rattling arguments predict self-organization towards states with small drive-induced fluctuations~\cite{Chvykov}. Readings of nonequilibrium selection in terms of entropy production assign dissipation the role of favouring some outcomes over others~\cite{Endres,Michaelian,KondepudiDeBari}, and the thermodynamics of switching between nonequilibrium steady states has been analysed directly~\cite{CookEndres}. Driven disordered matter encodes its loading history~\cite{Keim,Pashine,Gowen}, and in cyclically driven solids a transient rise in dissipation signals the formation of memory~\cite{ShohatLahini}. In plasticity, the work-hardening hypothesis makes the yield stress a function of the plastic work dissipated~\cite{Hill}. Positive-feedback loops make cell-fate decisions bistable and effectively irreversible~\cite{XiongFerrell}. Flashing and feedback ratchets change the potential in time, the latter according to the measured state~\cite{Reimann,CaoDinisParrondo}, and reinforced random processes give general criteria for when a reinforced choice persists~\cite{Pemantle}.

The model studied here combines a random choice with a specific reinforcement rule. A slow variable, the commitment $H$, accumulates the energy released when the selected structure recovers from perturbations, and stiffens the selected state [Eq.~\eqref{eq:commit}]. The rule is a constitutive closure, in the spirit of the work-hardening hypothesis~\cite{Hill}; it is not derived from a microscopic model, and it is a feedback-controlled potential whose full thermodynamic accounting, including the information processed by the feedback~\cite{SagawaUeda}, is not attempted here. Because the commitment starts at zero at the bifurcation, energy in this model can only reinforce a choice made by fluctuations, never make it: this is the model's working hypothesis, not one of its results. The results are the quantitative consequences of the closure for the growth of commitment and the statistics of reversal, together with the experiments that would test it.

\section{Model}\label{sec:model}
Let $x$ be the reduced coordinate along which a selected structure could be reversed (a wave-propagation direction, a domain orientation), with overdamped dynamics
\begin{equation}
\gamma\dot x=-\partial_x U_H(x)+\sqrt{2\gamma\kB T}\,\xi(t),\qquad \langle\xi(t)\xi(t')\rangle=\delta(t-t'),
\label{eq:langevin}
\end{equation}
\begin{equation}
U_H(x)=U(x)+\tfrac12\,a(H)\,(x-x^*)^2 ,\qquad a(H)=\lambda H .
\label{eq:Uh}
\end{equation}
Here $U$ is a symmetric double well, $x^*$ the branch selected at the bifurcation, $H$ (an energy) the \emph{commitment}, and $\lambda$ (inverse length squared) the coupling, so that the stiffness of the selected state is $\kappa(H)=\kappa_0+a(H)$. In Sec.~\ref{sec:sat} we replace the linear $a(H)$ by a saturating one. After selection the landscape is biased towards the chosen branch, $U_H(x_a)-U_H(x^*)=2a$ for the quartic well below, while the ensemble remains symmetric because $x^*$ is itself random.

\paragraph{Reversal.} A reversal is the event that $x$ reaches a fixed threshold $x_r$ on the side of the alternative branch, beyond the saddle of the unreinforced landscape. We take crossing $x_r$ to trigger an irreversible reorganization of the structure (a change of wave direction, a fracture), so that a crossing is a reversal whether or not an alternative minimum still exists. Before lock-in (Sec.~\ref{sec:results}) a crossing is an escape into the alternative basin; after lock-in it is a rare threshold crossing in a single well. The same event is used in all calculations and simulations.

\paragraph{Recovery events and their energy.} The structure is perturbed at discrete times. Perturbation $n$ brings it from $x_n^{\mathrm b}$ to $x_n^-$, doing work $W_n=U_H(x_n^-)-U_H(x_n^{\mathrm b})$, after which it recovers at fixed $H$ to $x_n^+$. Following stochastic energetics~\cite{Sekimoto,Seifert}, the heat released during recovery is $\dd Q=-\partial_xU_H\circ\dd x$, which for the conservative dynamics~\eqref{eq:langevin} at fixed $H$ is a difference of state functions,
\begin{equation}
q_n=U_H(x_n^-)-U_H(x_n^+).
\label{eq:qn}
\end{equation}
The model therefore applies to systems with conservative dynamics between perturbations, such as mechanical structures that are loaded and released; it has no housekeeping heat~\cite{OonoPaniconi}. In a driven steady state the excess heat would replace $q_n$, and heat and absorbed work would differ (Sec.~\ref{sec:exp}).

\paragraph{Postulate.} Commitment starts at $H=0$ at the bifurcation and grows by the energy released in each completed recovery,
\begin{equation}
H_{n+1}=H_n+q_n^+,\qquad q_n^+=\max(q_n,0).
\label{eq:commit}
\end{equation}
This is a constitutive closure. Its linearity in $q$ is an assumption, not the leading term of an expansion, since $q_n$ is several $\kB T$ in the regimes studied. The positive part makes commitment non-decreasing, so that a structure that is never disturbed never commits. This property is built into Eq.~\eqref{eq:commit}; it is not a finding about driven matter. Reinforcement is removed only by an explicit relaxation (Sec.~\ref{sec:sat}). The closure has one coupling, $\lambda$, which must be measured. Because $\lambda$ multiplies an energy rather than an energy in units of $\kB T$, the reinforcement per recovery stays finite as $T\to0$.

\paragraph{Energy balance.} Between recoveries, the update~\eqref{eq:commit} raises the energy at fixed $x$ by the stiffening work $\Delta W^{\mathrm{st}}_n=\tfrac12\lambda q_n^+(x_n^+-x^*)^2$, supplied by the reinforcing mechanism. Summing over cycles, with $q_n^-=\max(-q_n,0)$,
\begin{equation}
H_N-H_0=\sum_n q_n^+=\sum_n W_n+\sum_n\Delta W^{\mathrm{st}}_n-\Delta U+\sum_n q_n^- ,
\label{eq:balance}
\end{equation}
where $\Delta U$ is the change of $U_H$ between the first and last state. For a harmonic well and $q_n>0$, $\langle\Delta W^{\mathrm{st}}_n\rangle=(\lambda\kB T/2\kappa)(A-\tfrac32\kB T)$, with $A$ the injected energy defined below, a fraction of order $\lambda\kB T/2\kappa$ of the heat. The clipped part $\sum q_n^-$ is negligible only when perturbations inject several $\kB T$; for weak perturbations it dominates (Sec.~\ref{sec:results}). Within these limits the energy that feeds commitment is the work absorbed from the perturbing agent. The commitment can then be measured mechanically, as the area of the force--displacement loop of each cycle. By the same token, heat and absorbed work cannot be distinguished within the model: the adjective ``energy-fed'' in the title is chosen for that reason.

\paragraph{Reduced units.} We use $U(x)=x^4/4-x^2/2$ and $\gamma=1$, with energy and length scales absorbed into the reduced variables. Then $x^*=+1$, $\kappa_0=2$, the saddle is at $x_s=0$ with $d\equiv|x_s-x^*|=1$, the bare barrier is $\Delta U_0=1/4$, and $x_r=-1/2$. Results stated for the quartic well are exact for it and do not require small displacements; harmonic estimates are marked as such.

\section{Results}\label{sec:results}
\paragraph{Growth of commitment and control mode.} A perturbation injects the energy $A_n=U_H(x_n^-)-U_H(x^*)$. If the end of each recovery is equilibrated in the selected well, the energy above the minimum, $E_n=U_H(x_n^+)-U_H(x^*)$, is distributed for a harmonic well as $\kB T$ times a $\chi^2_1/2$ variable, independently of $H$. Averaging Eq.~\eqref{eq:commit} over $E_n$ then gives the approximate closure
\begin{equation}
\begin{gathered}
H_{n+1}\simeq H_n+\kB T\,g\!\left(\frac{A(H_n)}{\kB T}\right),\\
g(\alpha)=\Bigl(\alpha-\tfrac12\Bigr)\mathrm{erf}\sqrt\alpha+\sqrt{\tfrac{\alpha}{\pi}}\,e^{-\alpha},
\end{gathered}
\label{eq:closure}
\end{equation}
which we use as a stochastic map for each realization, sampling $E_n$ and reweighting by survival, not as an equation for the ensemble mean, since $g$ is convex and survival reweights the distribution of $H$. For $\alpha\gg1$, $g\to\alpha-\tfrac12$; at $A=\kB T/2$ the unclipped rule would give zero growth while Eq.~\eqref{eq:closure} gives $0.24\,\kB T$, and for $\alpha\ll1$, $g\simeq(4/3\sqrt\pi)\,\alpha^{3/2}$. The growth law depends on how the structure is perturbed:
\begin{itemize}
\item \emph{Displacement control}, $x_n^-=x^*-\delta$: for the quartic well exactly $A=Q_0+\tfrac12\lambda\delta^2H$, with $Q_0=\delta^2(2-\delta)^2/4$. For $Q_0\gg\kB T$,
\begin{equation}
H_{n+1}=(1+c)H_n+b,\qquad \langle q_n\rangle=b\,(1+c)^n,
\label{eq:hn}
\end{equation}
with $c=\lambda\delta^2/2$ and $b=Q_0-\kB T/2$. The heat per recovery and the stiffness increment $\lambda\langle q_n\rangle$ grow geometrically; the stiffness $\kappa_n=\kappa_0+(\lambda b/c)[(1+c)^n-1]$ is geometric only if $\lambda b/c\simeq\kappa_0$, i.e.\ for small harmonic displacements.
\item \emph{Energy control}, $A=E_0$ fixed: $H$ grows linearly, by $\kB T\,g(E_0/\kB T)$ per recovery.
\item \emph{Force control}, the structure held at the static balance point under a force $F$ and released: in the harmonic approximation $A\simeq F^2/2\kappa$, and while $A\gg\kB T$, $\kappa_n^2\simeq\kappa_0^2+\lambda F^2n$. Stiffening lowers $A$ until it is of order $\kB T$, after which Eq.~\eqref{eq:closure} gives the slower law $\kappa\propto n^{2/5}$.
\end{itemize}
The work absorbed per cycle therefore rises with commitment under displacement control, stays fixed under energy control, and falls under force control. In the last case the committed state absorbs less work from the drive the more committed it is. This is the same direction as the drive-specific selection of low-absorption states in multistable networks~\cite{Kedia}, which is reached by a different route; under displacement control the direction is opposite. Acceleration of commitment is a property of displacement control, not of the closure.

\paragraph{Reversal.} For the quartic well, with $a=\lambda H$ and $s=\sqrt{1-4a}$, the barrier and curvatures are exact (Appendix~\ref{app:quartic}):
\begin{equation}
\begin{gathered}
\Delta U=\frac{(3-s)^3(1+s)}{64},\\
\kappa=2+a,\qquad \mu\equiv|U_H''(x_s)|=\frac{s(3-s)}2 ,
\end{gathered}
\label{eq:exact}
\end{equation}
and Kramers' formula~\cite{Kramers,Hanggi} gives
\begin{equation}
\ln\frac{k(H)}{k(0)}=-\frac{\Delta U(H)-\Delta U_0}{\kB T}+\frac12\ln\frac{\kappa(H)\mu(H)}{\kappa_0\mu_0}.
\label{eq:law}
\end{equation}
For any smooth bistable potential the barrier grows to first order as $\tfrac12\lambda d^2H$, because $x^*$ and $x_s$ are stationary points of $U$, so the local slope is $\dd\ln k/\dd H|_0=-\lambda d^2/2\kB T+\tfrac12(\lambda/\kappa_0+\mu'_0/\mu_0)$, equal to $-\lambda[1/(2\kB T)+1/4]$ for the quartic well. This linearization holds only for $a\ll\kB T$; beyond it the $a^2/2\kB T$ term of the barrier dominates the prefactor term, and Eq.~\eqref{eq:law} must be used. Kramers' formula fails near the saddle-node at $s=0$.

Kramers' rate describes activated escape from a quasi-stationary state. A perturbation that brings the structure close to the saddle is followed instead by a transient, and the reversal probability within a recovery window $T_w$ is computed exactly from the survival function $S_H(x,t)$, the probability of not having reached $x_r$ by time $t$ from $x$:
\begin{equation}
\begin{gathered}
\partial_tS_H=\mathcal L_HS_H,\qquad S_H(x_r,t)=0,\qquad S_H(x,0)=1,\\
p(H)=1-S_H(x^-,T_w),
\end{gathered}
\label{eq:window}
\end{equation}
with $\mathcal L_H=-\gamma^{-1}U_H'\partial_x+\gamma^{-1}\kB T\partial_x^2$ and a reflecting boundary far from $x^*$. We solve Eq.~\eqref{eq:window} numerically (Appendix~\ref{app:num}). For $T_w\to\infty$ its transient part reduces to the splitting probability
\begin{equation}
\pi(H)=\frac{\int_{x^-}^{x^*}e^{U_H/\kB T}\dd x}{\int_{x_r}^{x^*}e^{U_H/\kB T}\dd x},
\label{eq:committor}
\end{equation}
which we use only for bounds.

\paragraph{Structural lock-in.}
\begin{proposition}
In the quartic model the alternative minimum and the saddle merge and disappear at $\lambda H_c=1/4$; for $H>H_c$ the selected state is the only minimum of $U_H$. Under displacement control with $\delta<x^*-x_r$, for every $\epsilon>0$ the splitting probability~\eqref{eq:committor} satisfies
\[
\ln\pi(H)\le C_\epsilon-\frac{\lambda H}{2\kB T}\bigl[(x^*-x_r-\epsilon)^2-\delta^2\bigr].
\]
\end{proposition}
\noindent Lock-in is a structural property of the landscape: it removes the alternative stable state, but crossings of $x_r$ remain possible at $T>0$ because the potential has no infinite wall. For the full window probability~\eqref{eq:window} we have not proven the corresponding bound; numerically, $\ln p$ decreases linearly in $\lambda H$ with slope $-23.5$ for $0.4\le\lambda H\le1.2$ ($p$ down to $10^{-9}$), against $-24.0$ from the splitting bound (Fig.~\ref{fig3}(b)). Under displacement control the deterministic recursion~\eqref{eq:hn} reaches $H_c$ after
\begin{equation}
n_c=\Bigl\lceil\frac{\ln(1+cH_c/b)}{\ln(1+c)}\Bigr\rceil
\label{eq:nc}
\end{equation}
recoveries; the number of recoveries at which an individual realization crosses $H_c$ is random.

\paragraph{Retention under unbounded reinforcement.} Let $P_n$ be the probability that perturbation $n$ reverses a structure that survived the previous ones, and $S_j=\prod_{n=1}^{j}(1-P_n)$, $S_0=1$.
\begin{proposition}
Let $\tilde H_n$ be the commitment chain driven by the recovery law conditioned on survival (Appendix~\ref{app:proof}). Assume (A1) every perturbation injects at least $Q_0>0$; (A2) there is $\rho>0$ such that, given the past, $\Pr(E_n\le Q_0/2)\ge\rho$; and (A3) $p(H)\le Ce^{-\alpha H}$ for some $\alpha>0$, and $p(H)\le p_{\max}<1$. Then $\tilde H_n\ge\rho Q_0n/4$ for all large $n$ almost surely, and $S_\infty>0$.
\end{proposition}
\noindent The proof (Appendix~\ref{app:proof}) uses only $q_n^+\ge(Q_0-E_n)^+$ and a strong law for bounded martingale increments. The assumptions carry much of the content. Linear growth matters: if commitment grew only as $\ln n$, hazards $p\propto e^{-\alpha H}$ could fail to be summable and $S_\infty=0$. (A1) holds for displacement and energy control but not for force control, where $A_n$ decreases. (A2) is a statement about local equilibrium in the selected well, which we check numerically but do not prove. (A3) holds for the splitting part by Proposition~1 and is verified numerically for the window probability. Proposition~2 states sufficient conditions, not necessary ones, and it requires reinforcement to continue without bound.

\paragraph{Saturation and forgetting.}\label{sec:sat} Unbounded stiffening is not physical, and any saturation changes the conclusion. With $a(H)=\lambda H_s(1-e^{-H/H_s})$, the hazard is bounded below by $p_s=p(a\to\lambda H_s)>0$, so $S_N\le C(1-p_s)^N\to0$: every choice is eventually lost, after a mean of about $1/p_s$ perturbations once the reinforcement has saturated. Because $\ln p$ decreases linearly in $a$ (Proposition~1), this lifetime grows exponentially with the saturation level $\lambda H_s$; structural lock-in, $\lambda H_s>1/4$, makes it long but not infinite. The same holds if reinforcement relaxes between perturbations, $H\to\phi H$ with $\phi=e^{-T_p/\tau_m}$, and $\phi(1+c)<1$. Then $q_n^+\le A_n=Q_0+cH_n$ gives the pathwise bound $H_n\le M=\phi Q_0/[1-\phi(1+c)]$, hence $S_N\le(1-p(M))^N$. For the deterministic recursion the fixed point is $H^*=\phi b/[1-\phi(1+c)]$, and structural lock-in at the times of perturbation requires $H^*>H_c$, i.e.
\begin{equation}
T_p<T_p^*=\tau_m\ln\Bigl(1+c+\frac{b}{H_c}\Bigr),
\label{eq:freq}
\end{equation}
with strict inequality. If $\phi(1+c)\ge1$ the recursion grows without bound despite the relaxation. ``Forever'' in Proposition~2 therefore means under continued perturbation with unbounded reinforcement. If perturbations stop at finite $H$ and thermal noise remains, a choice is eventually lost.

\begin{figure*}[t]
\centering
\includegraphics[width=\textwidth]{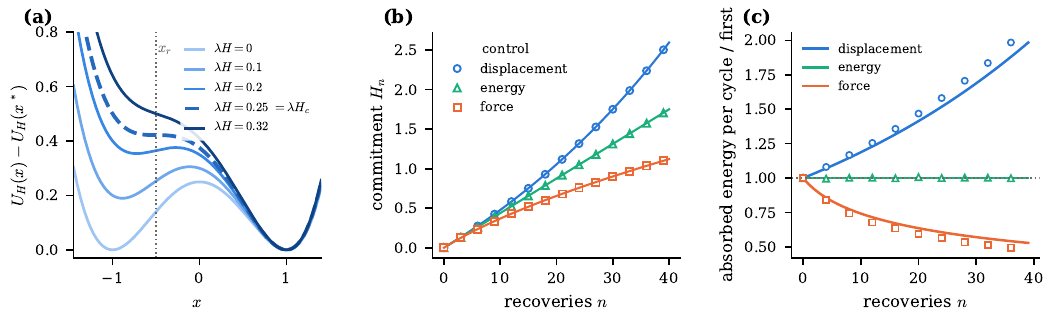}
\caption{\label{fig1}Landscape and control mode. (a) Landscape $U_H$ of the quartic model; the alternative minimum and the saddle merge at $\lambda H_c=1/4$ (dashed); the irreversibility threshold is $x_r=-1/2$ (dotted). (b) Commitment under displacement, energy and force control with a matched first pulse ($\kB T=0.01$, $\lambda=0.6$): simulation (symbols) and closure~\eqref{eq:closure} (lines). (c) Energy injected per cycle (lines) and mean recovery heat (symbols), relative to the first cycle: rising under displacement control, constant under energy control, falling under force control.}
\end{figure*}

\begin{figure*}[t]
\centering
\includegraphics[width=\textwidth]{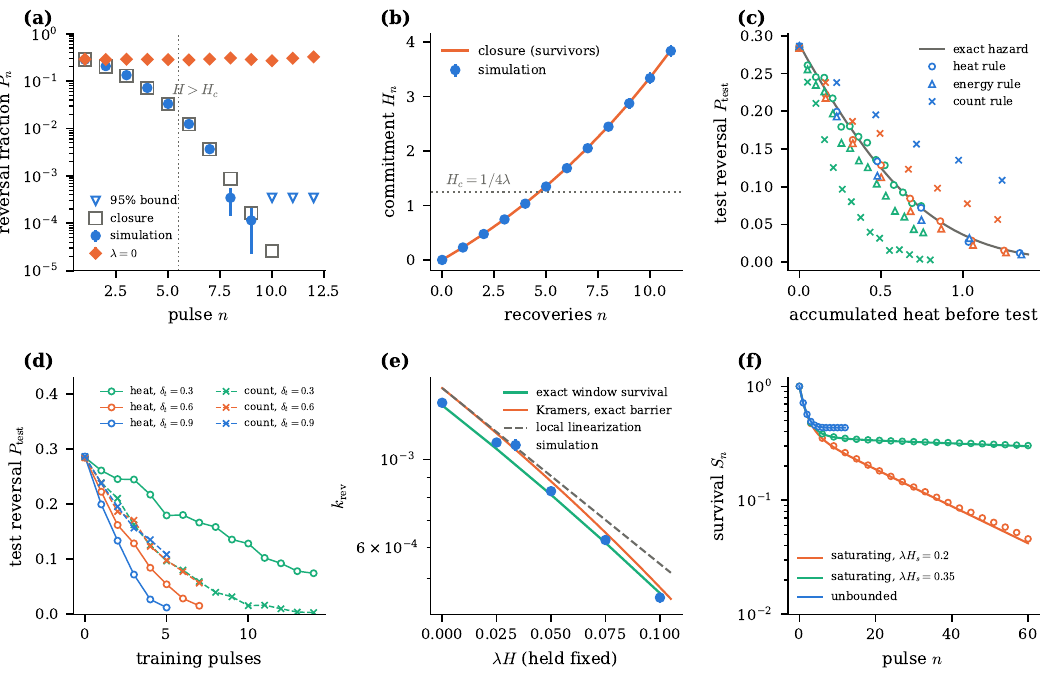}
\caption{\label{fig2}Numerical verification ($\kB T=0.03$, $\delta=0.9$, $\lambda=0.2$ unless stated). (a) Reversal fraction per identical pulse with the ratchet and without it ($\lambda=0$), binomial errors; open triangles, $95\%$ upper bound where no reversal occurred; open squares, closure with the exact window hazard~\eqref{eq:window}, no fit. (b) Mean commitment of survivors, simulation ($\pm$ s.d.) and closure; dotted line, $H_c$. (c) Test reversal ($\delta=0.9$) after training at amplitudes $0.3$ (green), $0.6$ (orange) and $0.9$ (blue) under three reinforcement rules (heat, injected energy, pulse count), against the accumulated recovery heat measured in each run; line, exact hazard. (d) The same test against the number of training pulses, heat and count rules. (e) Spontaneous reversal rate at fixed $\lambda H$ ($\kB T=0.05$): simulation, exact window survival, Kramers' formula~\eqref{eq:law}, and its local linearization. (f) Survival with unbounded and with saturating reinforcement: simulation (symbols) and closure (lines).}
\end{figure*}

\section{Numerical verification}\label{sec:num}
The simulations below integrate Eq.~\eqref{eq:langevin} and apply Eq.~\eqref{eq:commit} to the measured recovery heat. They verify the closure~\eqref{eq:closure}, the first-passage calculations and the propositions' assumptions within the model; they cannot test whether any material obeys Eq.~\eqref{eq:commit}. We used the Heun scheme at time step $2\times10^{-3}$, accumulated the heat with the Stratonovich midpoint rule (which here only checks the integration, since $q_n$ is a difference of state functions), equilibrated each realization in the selected well, and counted a reversal when $x$ reached $x_r=-1/2$ within a recovery window $T_w=8$ (the relaxation time is $\approx0.5$). Reversed realizations are removed, and $P_n$ is the reversed fraction of the survivors. Predictions use Eq.~\eqref{eq:closure} as a stochastic map with the exact window hazard~\eqref{eq:window}. Event counts for every pulse are provided with the code.

\emph{Base run.} With $2\times10^4$ realizations, $\kB T=0.03$, $\delta=0.9$ and $\lambda=0.2$, so that $c=0.081$, $b=7.67\,\kB T$ and $H_c=1.25$, the reversal fraction without reinforcement is constant, $P=0.290\pm0.007$ (s.d. over the first eight pulses) [Fig.~\ref{fig2}(a)]. With reinforcement the same pulses give $P_n=0.284$, $0.207$, $0.134$, $0.072$, $0.033$, $0.012$, $3.7\times10^{-3}$, $3.5\times10^{-4}$ and $1.2\times10^{-4}$, and no reversal among the $8680$ survivors of the last three pulses ($95\%$ upper bound $3.5\times10^{-4}$ per pulse). The closure predicts $0.288$, $0.204$, $0.130$, $0.072$, $0.033$, $0.012$, $3.7\times10^{-3}$, $8.7\times10^{-4}$ and $1.7\times10^{-4}$, within $1.7$ binomial standard errors at every pulse. The survival after twelve pulses is $0.434$, against $0.435$. The mean commitment crosses $H_c$ at the fifth recovery, as Eq.~\eqref{eq:nc} gives, when $93\%$ of the survivors have crossed it; threshold crossings then continue at the small rate predicted. The mean heat per recovery grows by a factor between $1.080$ and $1.084$ per cycle, against $1+c=1.081$, and equals the mean injected work within $0.5\%$; the stiffening work is $0.11$--$0.16\%$ of the heat, consistent with the harmonic estimate. The fraction of survivors with $E_n\le Q_0/2$, assumption (A2), is at least $0.991$ at every pulse. Halving the time step changes $H_n$ by less than $0.4\%$ and $P_n$ within binomial error.

\emph{Control mode.} With $\kB T=0.01$, $\lambda=0.6$ and a force $F=0.33$ whose static displacement $\delta_0=0.253$ injects $E_0=4.9\,\kB T$, we perturbed $4000$ realizations forty times at fixed displacement $\delta_0$, at fixed injected energy $E_0$, and at fixed force $F$. After forty recoveries the commitment is $2.60$, $1.75$ and $1.12$, respectively, within $0.3\%$ of the closure [Fig.~\ref{fig1}(b)], and the injected energy per cycle has changed by factors $1.98$, $1.00$ and $0.53$ [Fig.~\ref{fig1}(c)]. No reversals occur at these amplitudes.

\emph{What the reversal statistics can discriminate.} We trained three ensembles ($4000$ realizations each) at amplitudes $\delta_t=0.3$, $0.6$ and $0.9$ and gave copies a test pulse with $\delta=0.9$ after each training pulse, under three reinforcement rules: the heat rule~\eqref{eq:commit}, an injected-energy rule $H\mathrel{+}=A_n$, and a count rule $H\mathrel{+}=0.12$ per pulse. Against the recovery heat accumulated in each run, the heat rule gives a single curve for all amplitudes, on the exact-hazard prediction; the energy rule nearly does, with deviations at the weakest amplitude, where the injected energy exceeds the released heat by the residual thermal energy; the count rule does not [Fig.~\ref{fig2}(c)]. Against the number of training pulses the count rule collapses and the heat rule separates by amplitude [Fig.~\ref{fig2}(d)]. Under a single protocol of identical pulses all three rules produce similar ageing. Training at several amplitudes therefore separates energy-fed reinforcement from counting. It does not separate heat from absorbed work, which are equal on average by Eq.~\eqref{eq:balance}.

\emph{Spontaneous reversal.} Holding $\lambda H$ fixed at $\kB T=0.05$ ($\Delta U_0/\kB T=5$), we measured the fraction of realizations started at $x^*$ that reach $x_r$ within $\tau=100$, with $k=-\ln(1-p)/\tau$ [Fig.~\ref{fig2}(e)]. The rates agree with the exact window survival~\eqref{eq:window} within $1.7$ standard errors at every point. The weighted slope $\dd\ln k/\dd(\lambda H)=-11.1\pm0.3$ compares with $-10.9$ from the exact survival, $-11.1$ from the exact barrier alone, $-11.6$ from Kramers' formula~\eqref{eq:law} with its prefactor, and $-10.25$ from the local linearization. The data do not resolve the prefactor term; the linearization fails here because $a$ reaches $2\kB T$.

\emph{Saturation.} With $a(H)=\lambda H_s(1-e^{-H/H_s})$ and $10^4$ realizations over sixty pulses, the reversal fraction levels off at $0.035$ for $\lambda H_s=0.2$, against $p_s=0.037$, and at $3.0\times10^{-3}$ for $\lambda H_s=0.35$, beyond lock-in, against $p_s=2.9\times10^{-3}$. The survival after sixty pulses is $0.046$ (closure $0.042$) and $0.302$ (closure $0.298$) [Fig.~\ref{fig2}(f)], and keeps falling: mean lifetimes after saturation are about $27$ and $350$ pulses.

\emph{Forgetting.} With relaxation, Eq.~\eqref{eq:freq} gives $\phi^*=0.791$. For $\phi=0.75$, $H^*=0.91<H_c$ and the reversal fraction levels off near $0.046$, against $p(H^*)=0.047$; $7.9\%$ of choices survive thirty pulses (closure $7.8\%$). For $\phi=0.85$, $H^*=2.41>H_c$; reversal becomes rare but not zero, and $30.1\%$ survive thirty pulses (closure $30.6\%$).

\emph{Robustness.} Table~\ref{tab} varies $\lambda$, $\kB T$ and $\delta$. In all eight sets the survival after ten pulses agrees with the closure within $0.009$. The lock-in pulse of the recursion coincides with the pulse at which half of the survivors have crossed $H_c$ whenever the run reaches it.

\begin{table*}[t]
\centering\small
\caption{\label{tab}Parameter sensitivity ($10^4$ realizations, ten pulses). $c=\lambda\delta^2/2$; $n_c$: recoveries to lock-in from the deterministic recursion, and pulse at which half of the survivors have crossed $H_c$ in the simulation (``$>9$'': not reached); $P^{(0)}$: reversal fraction without reinforcement (mean $\pm$ s.d. over pulses); $P_1$, $P_5$, $S_{10}$: first- and fifth-pulse reversal fractions and survival after ten pulses with reinforcement, simulation (closure with exact hazard).}
\begin{tabular}{@{}ccccccccc@{}}
\toprule
$\lambda$ & $\kB T$ & $\delta$ & $c$ & $n_c$ (recursion/sim.) & $P^{(0)}$ & $P_1$ & $P_5$ & $S_{10}$\\
\midrule
0.2 & 0.03 & 0.9 & 0.081 & 5 / 5 & $0.289\pm0.006$ & 0.296 (0.288) & 0.030 (0.033) & 0.426 (0.435)\\
0.1 & 0.03 & 0.9 & 0.041 & 10 / $>9$ & $0.286\pm0.014$ & 0.286 (0.288) & 0.130 (0.127) & 0.245 (0.239)\\
0.4 & 0.03 & 0.9 & 0.162 & 3 / 3 & $0.284\pm0.010$ & 0.283 (0.288) & $1.7\times10^{-4}$ ($2.9\times10^{-4}$) & 0.588 (0.589)\\
0.2 & 0.025 & 0.9 & 0.081 & 5 / 5 & $0.267\pm0.006$ & 0.271 (0.268) & 0.023 (0.020) & 0.494 (0.495)\\
0.2 & 0.04 & 0.9 & 0.081 & 5 / 5 & $0.317\pm0.011$ & 0.319 (0.320) & 0.064 (0.063) & 0.328 (0.336)\\
0.2 & 0.03 & 0.95 & 0.090 & 5 / 5 & $0.380\pm0.017$ & 0.393 (0.391) & 0.055 (0.052) & 0.290 (0.287)\\
0.2 & 0.03 & 0.8 & 0.064 & 6 / 6 & $0.134\pm0.007$ & 0.132 (0.133) & 0.014 (0.013) & 0.714 (0.712)\\
0.2 & 0.03 & 0.7 & 0.049 & 6 / 6 & $0.051\pm0.003$ & 0.052 (0.050) & 0.005 (0.005) & 0.884 (0.884)\\
\bottomrule
\end{tabular}
\end{table*}

\section{Experimental tests}\label{sec:exp}
A test requires a structure selected at an instability that can be perturbed repeatedly, with independent measurements of (i) the fraction reversed at each perturbation, (ii) the energy absorbed in each cycle, from the force--displacement loop where a mechanical handle exists, and (iii) the stiffness or recovery rate. Within its scope the model predicts:

\emph{Stiffening per recovery.} Each stiffness increment equals $\lambda$ times the energy absorbed in that cycle. Under displacement control the absorbed energy grows by a constant factor $1+c$ per cycle; under energy or force control the growth is linear or decelerating. The first cycles calibrate $\lambda$; a second protocol must then be predicted without refitting.

\emph{Separation from counting.} Training at several amplitudes and comparing at equal accumulated absorbed energy, the reversal statistics coincide; compared at equal pulse number they do not [Fig.~\ref{fig2}(c,d)]. This separates energy-fed reinforcement from reinforcement that counts perturbations; a single protocol of identical pulses does not.

\emph{Sign of the absorption trend.} Committed structures should absorb more work per cycle under displacement control and less under force control (Fig.~\ref{fig1}(c)).

\emph{Heat versus work.} In this model heat and absorbed work coincide. They differ in driven steady states with nonzero housekeeping heat, or when part of the absorbed energy is stored in long-lived defects. There the closure~\eqref{eq:commit}, with excess heat in place of $q_n$, predicts $\Delta\kappa_n=\lambda q_n$ rather than $\lambda W_n$, and that is the experiment that would test whether heat, rather than absorbed work, is the currency. Time-irreversibility estimators~\cite{Roldan} give lower bounds on entropy production, not the heat of each recovery, so they can support but not replace such measurements.

A reversal fraction that stays flat falsifies the closure only if $\lambda$ has been measured independently and the number of realizations suffices to resolve the predicted decrease. Candidate platforms are trainable mechanical networks and elastomer metamaterials~\cite{Pashine,Gowen} and cyclically driven crumpled sheets~\cite{ShohatLahini}, in which absorbed energy and stiffness are measured cycle by cycle. We have no estimate of $\lambda$ for any real system.

\section{Relation to previous work}\label{sec:prev}
The model is a feedback-controlled potential, and we make no claim about its consistency with fluctuation relations for driven systems~\cite{England,Perunov} beyond energy conservation. Those relations hold for a fixed driving protocol; here the potential is updated from measured heat, so a fluctuation-theorem accounting would have to include the reinforcing mechanism and the information it uses~\cite{SagawaUeda}. The reversal probability computed here, a crossing of $x_r$ at fixed reinforcement, is not the reverse-protocol probability that enters those relations. In entropy-production readings of selection~\cite{Endres,Michaelian,KondepudiDeBari}, dissipation favours some outcomes over others; in this model energy enters only after the choice, by assumption. A test between the two pictures requires branches that differ in dissipation within one dynamics, which the present model does not contain. Drive-specific selection in multistable networks~\cite{Kedia,GoldEngland} and low-rattling arguments~\cite{Chvykov} favour states that absorb little work from the drive or fluctuate little under it; under force control the closure moves committed states in the same direction, by stiffening rather than by selecting among configurations. The closure is closest in form to the work-hardening hypothesis of plasticity~\cite{Hill}, transposed from plastic work to the energy of elastic recoveries and applied to a bistable structure. The observation that dissipation signals memory formation in driven disordered solids~\cite{ShohatLahini} is a correlation; the closure postulates, rather than derives, a reinforcement law of that kind. Bistable memory modules built on positive feedback~\cite{XiongFerrell} are a biochemical analogue of commitment without a thermodynamic currency. The permanence of the record is an instance of the summability criterion for reinforced processes~\cite{Pemantle}, and lock-in by self-reinforcement echoes increasing returns~\cite{Arthur}. In adaptive transport networks such as \emph{Physarum}, reinforcement follows the steady flux~\cite{Tero,HuCai}; in this model steady operation does not commit, by construction of Eq.~\eqref{eq:commit}.

\section{Scope and limitations}\label{sec:limits}
Equation~\eqref{eq:commit} is a phenomenological closure: $\lambda$ must be measured, and the link between recovery energy and stiffness has not been derived from a microscopic model. The model is conservative between perturbations, one-dimensional, assumes local equilibrium at the end of each recovery, and its quantitative predictions are for the quartic well. That energy reinforces rather than selects is assumed. Proposition~2 rests on (A1)--(A3), of which (A2) and the activated part of (A3) are checked numerically, and it requires unbounded reinforcement; with saturation or relaxation, retention is finite. The simulations verify the analysis of the closure, not the closure. Nothing here involves replication or heredity, so the model is not a theory of biological selection.

The natural next step is a model in which $H$ is a physical slow coordinate with its own bath, driven by dissipation, so that Eq.~\eqref{eq:commit} can emerge or fail, its entropy production can be computed, and the assumption that energy reinforces rather than selects can be tested in a single dynamics that also contains the bifurcation, for instance with branches of unequal dissipation.

\section{Conclusion}
A one-parameter closure, in which the energy released in each recovery stiffens a state selected by fluctuations, yields a control-dependent growth of commitment, a reversal rate that falls exponentially with the absorbed energy, a structural lock-in threshold, and explicit conditions under which a random choice is retained forever or, with saturating or relaxing reinforcement, for a time exponential in the saturation level. In this model heat and absorbed work coincide, so its predictions can be tested mechanically; whether heat itself is the currency can be decided only where the two differ.

\paragraph{Funding and competing interests.} This work received no external funding. The author declares no competing interests.

\paragraph{Use of AI tools.} The central hypothesis is the author's. An AI assistant (Claude, Anthropic) was used to help draft the text, check the derivations, and write the simulation code. The author reviewed all content and takes full responsibility for it.

\paragraph{Data and code availability.} The Python code that generates every figure, table and quoted number, including per-pulse event counts, is provided as ancillary files with this preprint.

\appendix
\section{Quartic landscape}\label{app:quartic}
With $a=\lambda H$, $U_H'(x)=(x-1)(x^2+x+a)$, so $x^*=1$ and, for $a<1/4$, the saddle and alternative minimum are $x_{s,a}=(-1\pm s)/2$ with $s=\sqrt{1-4a}$. Substituting $a=(1-s^2)/4$ gives $U_H(x_s)-U_H(1)=(3-s)^3(1+s)/64=\frac14+\frac a2+\frac{a^2}2+O(a^3)$. The curvatures are $U_H''(1)=2+a$ and $U_H''(x_s)=-s(3-s)/2$, so $\frac{\dd}{\dd a}\ln\sqrt{\kappa\mu}\,|_{a=0}=-\frac14$. For a displacement to $1-\delta$, $U_H(1-\delta)-U_H(1)=\frac14\delta^2[2a+(2-\delta)^2]$, which gives $Q_0$ and $c$. For $x\in(x_r,\infty)$ and $a\ge0$, $U_H(x)\ge U_H(1)$, so $E_n\ge0$ for survivors. All expressions were checked symbolically.

\section{Proofs}\label{app:proof}
\emph{Proposition 1, bound.} The numerator of Eq.~\eqref{eq:committor} is at most $\delta\,e^{\max_{[x^-,x^*]}U_H/\kB T}$ and the denominator at least $\epsilon\,e^{\min_{[x_r,x_r+\epsilon]}U_H/\kB T}$. For large $H$ the quadratic term dominates both, giving the stated bound; since $\pi\le1$ and is continuous, it holds for all $H$ with a suitable $C_\epsilon$.

\emph{Proposition 2.} At state $H$ the structure is reversed with probability $p(H)$ and otherwise moves to $H+q^+$ with the law of $q^+$ conditioned on survival; let $\tilde H_n$ be the chain driven by this conditioned law without removal. By induction on $N$ and the Markov property, $S_N=\mathbb E\prod_{n<N}[1-p(\tilde H_n)]$, the Feynman--Kac representation of killing. Since $E_n\ge0$ and $A_n\ge Q_0$, $q_n^+\ge(Q_0-E_n)^+\ge\frac{Q_0}2B_n$ with $B_n=\mathbb 1\{E_n\le Q_0/2\}$. By (A2), $\mathbb E[B_n\mid\text{past}]\ge\rho$; the centred sums of $B_n$ form a martingale with bounded increments, so $\frac1n\sum_{k<n}(B_k-\mathbb E[B_k\mid\text{past}])\to0$ almost surely. Hence $\tilde H_n\ge\rho Q_0n/4$ for $n\ge n_0(\omega)$, and by (A3) $\sum_np(\tilde H_n)<\infty$ almost surely. With $p\le p_{\max}<1$, $\prod_n[1-p(\tilde H_n)]\ge\exp[-\sum_np(\tilde H_n)/(1-p_{\max})]>0$, so $S_\infty>0$ by monotone convergence.

\emph{Saturation and forgetting.} With saturation, $a(H)<\lambda H_s$ and $p$ decreasing in $a$ give $p(\tilde H_n)\ge p_s$; with relaxation and $\phi(1+c)<1$, $\tilde H_n\le M$ and $p(\tilde H_n)\ge p(M)$. In both cases $S_N\to0$ geometrically. That $p$ decreases in $a$ is verified numerically for the parameters used.

\section{Selection at the instability}\label{app:ramp}
\begin{figure}[t]
\centering
\includegraphics[width=\columnwidth]{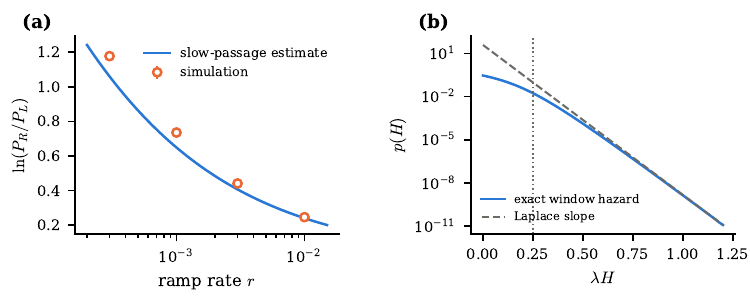}
\caption{\label{fig3}(a) Selection odds versus ramp rate at identical final control parameters, simulation of Eq.~\eqref{eq:twomode} and Eq.~\eqref{eq:ramp}. (b) Exact window hazard $p$ versus $\lambda H$ ($\kB T=0.03$, $\delta=0.9$) and the slope of the splitting bound of Proposition~1; dotted line, lock-in.}
\end{figure}

Part of the working hypothesis is that the branch is chosen by fluctuations and by how the instability is crossed. As an illustration, separate from the reinforcement model, consider two global-mode amplitudes with winner-take-all coupling $g>1$~\cite{CrossHohenberg},
\begin{equation}
\dd A_j=\bigl[\varepsilon_jA_j-|A_j|^2A_j-g|A_k|^2A_j\bigr]\dd t+\sqrt{2D}\,\dd W_j ,
\label{eq:twomode}
\end{equation}
$j\neq k\in\{R,L\}$, independent complex noises with $\langle\dd W_j\dd W_j^*\rangle=\dd t$, reduced amplitude units, and $\varepsilon_{R,L}=rt\pm\Delta\varepsilon$. In the linear stage the variances are $V_\pm=2De^{rt^2\pm2\Delta\varepsilon t}\int_{-\infty}^te^{-rs^2\mp2\Delta\varepsilon s}\dd s$, so $V_+/V_-\simeq e^{4\Delta\varepsilon t}$; for exponentially distributed intensities $P_R/P_L\simeq V_+/V_-$, and freezing the winner at the delayed onset $\varepsilon^*$ gives the classical slow-passage estimate~\cite{MandelErneux,SwiftHohenbergAhlers,KondepudiNelson}
\begin{equation}
\ln\frac{P_R}{P_L}\simeq\frac{4\Delta\varepsilon\,\varepsilon^*}{r},\qquad \frac{\varepsilon^{*2}}{r}=\ln\frac{\varepsilon^*\sqrt r}{2D\sqrt\pi}.
\label{eq:ramp}
\end{equation}
The conversion of the linear advantage into a final choice is not derived. We integrated Eq.~\eqref{eq:twomode} ($2\times10^4$ realizations, $D=10^{-6}$, $\Delta\varepsilon=2\times10^{-3}$, $g=2$), ramping to the same final value $\varepsilon_f=0.5$ for every $r$ and then holding it with the noise switched off, so that no hopping between branches occurs after the ramp. The odds are $\ln(P_R/P_L)=0.25$, $0.44$, $0.74$ and $1.18$ ($\pm0.02$) for $r=10^{-2}$, $3\times10^{-3}$, $10^{-3}$ and $3\times10^{-4}$, within $14\%$ of Eq.~\eqref{eq:ramp} [Fig.~\ref{fig3}(a)]. The odds depend on the crossing rate at identical final control parameters, a classical result we include only to make the working hypothesis concrete; with noise maintained for times longer than the hopping time, the stationary odds of this variational model would depend on the final parameters alone.

\section{Numerical methods}\label{app:num}
Equation~\eqref{eq:window} was solved by Crank--Nicolson on $[x_r,2.5]$ with grid spacing $2\times10^{-3}$, time step $5\times10^{-3}$, absorbing boundary at $x_r$ and reflecting boundary at $x=2.5$; halving both changes $p$ by less than $10^{-5}$ at $\lambda H=0$, and the solver reproduces the exact splitting probability~\eqref{eq:committor} to $5\times10^{-4}$ in the long-window limit of the transient. Hazards were tabulated on a grid in $\lambda H$ and interpolated in $\ln p$; values below $10^{-11}$ are not resolved.

\begin{thebibliography}{99}\small
\bibitem{Prigogine} G. Nicolis and I. Prigogine, \emph{Self-Organization in Nonequilibrium Systems} (Wiley, New York, 1977).
\bibitem{CrossHohenberg} M. C. Cross and P. C. Hohenberg, Rev. Mod. Phys. \textbf{65}, 851 (1993).
\bibitem{Kondepudi} D. K. Kondepudi, R. J. Kaufman, and N. Singh, Science \textbf{250}, 975 (1990).
\bibitem{KondepudiDeBari} D. K. Kondepudi, B. De Bari, and J. A. Dixon, Entropy \textbf{22}, 1305 (2020).
\bibitem{DeBari} B. De Bari \emph{et al.}, BioSystems \textbf{239}, 105214 (2024).
\bibitem{PrigogineScience} I. Prigogine, Science \textbf{201}, 777 (1978).
\bibitem{Michaelian} K. Michaelian, Entropy \textbf{23}, 217 (2021).
\bibitem{England} J. L. England, J. Chem. Phys. \textbf{139}, 121923 (2013); Nat. Nanotechnol. \textbf{10}, 919 (2015).
\bibitem{Perunov} N. Perunov, R. A. Marsland, and J. L. England, Phys. Rev. X \textbf{6}, 021036 (2016).
\bibitem{Kedia} H. Kedia, D. Pan, J.-J. Slotine, and J. L. England, J. Chem. Phys. \textbf{159}, 214106 (2023).
\bibitem{GoldEngland} J. M. Gold and J. L. England, arXiv:1911.07216.
\bibitem{Chvykov} P. Chvykov \emph{et al.}, Science \textbf{371}, 90 (2021).
\bibitem{Endres} R. G. Endres, Sci. Rep. \textbf{7}, 14437 (2017).
\bibitem{CookEndres} J. Cook and R. G. Endres, arXiv:1908.07405.
\bibitem{Keim} N. C. Keim, J. D. Paulsen, Z. Zeravcic, S. Sastry, and S. R. Nagel, Rev. Mod. Phys. \textbf{91}, 035002 (2019).
\bibitem{Pashine} N. Pashine, D. Hexner, A. J. Liu, and S. R. Nagel, Sci. Adv. \textbf{5}, eaax4215 (2019).
\bibitem{Gowen} S. D. Gowen \emph{et al.}, Proc. Natl. Acad. Sci. U.S.A. \textbf{122}, e2504304122 (2025).
\bibitem{ShohatLahini} D. Shohat and Y. Lahini, Phys. Rev. Lett. \textbf{130}, 048202 (2023).
\bibitem{Hill} R. Hill, \emph{The Mathematical Theory of Plasticity} (Clarendon Press, Oxford, 1950).
\bibitem{XiongFerrell} W. Xiong and J. E. Ferrell, Nature \textbf{426}, 460 (2003).
\bibitem{Reimann} P. Reimann, Phys. Rep. \textbf{361}, 57 (2002).
\bibitem{CaoDinisParrondo} F. J. Cao, L. Dinis, and J. M. R. Parrondo, Phys. Rev. Lett. \textbf{93}, 040603 (2004).
\bibitem{Pemantle} R. Pemantle, Probab. Surv. \textbf{4}, 1 (2007).
\bibitem{SagawaUeda} T. Sagawa and M. Ueda, Phys. Rev. Lett. \textbf{104}, 090602 (2010).
\bibitem{Sekimoto} K. Sekimoto, Prog. Theor. Phys. Suppl. \textbf{130}, 17 (1998).
\bibitem{Seifert} U. Seifert, Rep. Prog. Phys. \textbf{75}, 126001 (2012).
\bibitem{OonoPaniconi} Y. Oono and M. Paniconi, Prog. Theor. Phys. Suppl. \textbf{130}, 29 (1998); T. Hatano and S.-i. Sasa, Phys. Rev. Lett. \textbf{86}, 3463 (2001).
\bibitem{Kramers} H. A. Kramers, Physica \textbf{7}, 284 (1940).
\bibitem{Hanggi} P. H\"anggi, P. Talkner, and M. Borkovec, Rev. Mod. Phys. \textbf{62}, 251 (1990).
\bibitem{Roldan} \'E. Rold\'an, J. Barral, P. Martin, J. M. R. Parrondo, and F. J\"ulicher, New J. Phys. \textbf{23} (2021), doi:10.1088/1367-2630/ac0f18.
\bibitem{Arthur} W. B. Arthur, Econ. J. \textbf{99}, 116 (1989).
\bibitem{Tero} A. Tero \emph{et al.}, Science \textbf{327}, 439 (2010).
\bibitem{HuCai} D. Hu and D. Cai, Phys. Rev. Lett. \textbf{111}, 138701 (2013).
\bibitem{MandelErneux} P. Mandel and T. Erneux, J. Stat. Phys. \textbf{48}, 1059 (1987); S. M. Baer, T. Erneux, and J. Rinzel, SIAM J. Appl. Math. \textbf{49}, 55 (1989).
\bibitem{SwiftHohenbergAhlers} J. B. Swift, P. C. Hohenberg, and G. Ahlers, Phys. Rev. A \textbf{43}, 6572 (1991).
\bibitem{KondepudiNelson} D. K. Kondepudi and G. W. Nelson, Nature \textbf{314}, 438 (1985).
\end{thebibliography}
\end{document}